\documentstyle[preprint,aps,prd]{revtex}
\tightenlines
\newcommand{\be}{\begin{equation}}
\newcommand{\ee}{\end{equation}}
\newcommand{\ba}{\begin{eqnarray}}
\newcommand{\ea}{\end{eqnarray}}
\begin{document}
\draft 
\title{Torsion and the Electromagnetic Field}
\author{V. C. de Andrade and J. G. Pereira}
\vskip 0.5cm
\address{Instituto de F\'{\i}sica Te\'orica\\
Universidade Estadual Paulista\\
Rua Pamplona 145\\
01405-900\, S\~ao Paulo \\ 
Brazil}
\maketitle
\begin{abstract}

In the framework of the teleparallel equivalent of general
relativity, we study the dynamics of a gravitationally coupled
electromagnetic field. It is shown that the electromagnetic field
is able not only to couple to torsion, but also, through its
energy--momentum tensor, to produce torsion. Furthermore, it is
shown that the coupling of the electromagnetic field  with torsion
preserves the local gauge invariance of Maxwell's theory.

\end{abstract}

\section{Introduction}

Since the early days of general relativity, theoretical speculations have 
discussed the necessity of including torsion in the description of the
gravitational interaction. On the other hand, as is widely known, in the
framework of Einstein--Cartan theory,\cite{ec} the electromagnetic field
cannot be coupled to torsion in order to preserve the local gauge
invariance.\cite{sabbata} Equivalently, one can say that, in the presence
of torsion, the requirement of gauge invariance precludes the existence of a
gravitational minimal coupling prescription for the electromagnetic
field.\cite{heyde} To circumvent this problem, it is usually supposed
that, in an Einstein--Cartan background, the electromagnetic field can
neither produce nor feel torsion. In other words, torsion is assumed to
be irrelevant to the Maxwell's equations.\cite{bdt} It should be remarked
that this
hypothesis is not valid at a microscopic level since, from a quantum
point of view, one may always expect an interaction between photons and
torsion.\cite{sabbata} The reason for this is that a photon,
perturbatively speaking, can virtually disintegrate into an
electron--positron pair, and as these particles are massive fermions
which couple to torsion, the photon must necessarily feel the presence of
torsion. Consequently, even not interacting directly with torsion, the
photon field does feel torsion through the virtual pair produced by the
vacuum polarization. However, as all macroscopic phenomena must necessarily
have an interpretation based on an average of microscopic phenomena, and taking
into account the strictly attractive character of gravitation which eliminates
the possibility of a vanishing average, the above result seems to lead to
a contradiction because no interaction is usually
supposed to exist at the macroscopic level.

Our only concern in this paper will be the macroscopic
long-range gravitational interaction. In this context, an important point
to be considered is the following: there exists no compelling experimental
reasons to include torsion,
{\it besides curvature}, in the description of the gravitational interaction.
Furthermore, in the framework of the teleparallel equivalent of general
relativity,\cite{hayshi} despite presenting quite different
characteristics, curvature and torsion provide each one a complete
description of the gravitational interaction. According to general
relativity, curvature is used to {\it geometrize} spacetime, and in this
way successfully describe the gravitational interaction. Teleparallelism,
on the other hand, attributes gravitation to torsion, but in this case
torsion accounts for gravitation not by geometrizing the interaction, but
by acting as a {\it force}.\cite{paper} According to this approach,
gravitation might
present two alternative descriptions in which the relevant dynamical
fields are respectively the curvature and torsion tensors. Consequently,
as the electromagnetic field is able to couple to curvature,
relying upon the alluded equivalence, we can conclude that it should also
be able to couple to torsion, even at the macroscopic level.

With the purpose of studying the interaction of Maxwell's field with
gravitation, the latter being described
by the teleparallel equivalent of general relativity, we assume the scene
of this work to be a
spacetime manifold on which a nontrivial tetrad field is defined. The
context will be that of a gauge theory for the translation
group,\cite{hn} in which the gravitational field appears as the nontrivial
part of the tetrad.\cite{paper} We will use the greek alphabet ($\mu$,
$\nu$, $\rho$,~$\cdots=1,2,3,4$) to denote tensor indices,  that is,
indices related to spacetime. The latin alphabet ($a$, $b$,
$c$,~$\cdots=1,2,3,4$) will be used to denote local Lorentz (or tangent
space) indices. Of course, being of the same kind, tensor and local
Lorentz indices can be changed into each other with the use of a
tetrad $h^{a} {}_{\mu}$, which satisfy
\be
h^{a}{}_{\mu} \; h_{a}{}^{\nu} = \delta_{\mu}{}^{\nu} \quad
; \quad h^{a}{}_{\mu} \; h_{b}{}^{\mu} =
\delta^{a}{}_{b} \; .
\label{orto}
\ee

A nontrivial tetrad field induces on spacetime both a metric and a
teleparallel structures. In section 2, by emphasizing the
role played by the connections, we make a study of these
geometric structures, and review the well known result establishing the
equivalence between general relativity and a gauge theory for the
translation group. In section 3, a minimal coupling prescription is
introduced in terms of the Fock--Ivanenko derivative operator, and it
is shown how it reduces to the usual covariant derivative of
general relativity. Then, for the sake of completeness, Maxwell's theory
is briefly described in the framework of general relativity. In section 4,
the dynamics of a gravitationally coupled electromagnetic field is
described in terms of the teleparallel structure induced in spacetime by
the presence of the gravitational field. In this context, a new coupling
prescription is introduced, which is a natural consequence of the alluded
equivalence between general relativity and a gauge theory for the translation
group. Then, we show
that, provided this coupling prescription is used,
torsion is found not to violate the local gauge symmetry of Maxwell's
theory. Furthermore, by considering a system formed by electromagnetic plus
gravitational fields, we show that, besides coupling to torsion, the
electromagnetic field can also produce torsion.
Finally, in section 5, we draw the conclusions of the paper.

\section{Riemannian and Teleparallel Descriptions of Gravitation}

Differently from what is usually done, in what follows we will separate
the notions of space and connections. From a formal point of view,
curvature and torsion are in fact properties of a connection.\cite{koba}
Strictly speaking, there is no such a thing as curvature or torsion of
spacetime, but only curvature or torsion of connections. This becomes
evident if we notice that different particles feel different connections,
and consequently show distinct trajectories in spacetime. In the specific
case of general relativity, it is worth mentioning, universality of
gravitation allows the Levi--Civita connection to be interpreted as
part of the spacetime definition as all particles and fields feel this
connection the
same. It seems far wiser, however, to take spacetime simply as a
manifold, and connections (with their curvatures and torsions) as
additional structures.

Curvature and torsion, therefore, will be considered as properties of
connections, and many different connections are allowed to exist on the
same space.\cite{livro} For example, denoting by $\eta_{a b}$ the metric
tensor of the tangent space, a nontrivial tetrad field can be used to
define the riemannian metric
\be
g_{\mu \nu} = \eta_{a b} \; h^a{}_\mu \; h^b{}_\nu \; ,
\label{gmn}
\ee
in terms of which the Levi--Civita connection 
\be
{\stackrel{\circ}{\Gamma}}{}^{\sigma}{}_{\mu \nu} = \frac{1}{2} 
g^{\sigma \rho} \left[ \partial_{\mu} g_{\rho \nu} + \partial_{\nu}
g_{\rho \mu} - \partial_{\rho} g_{\mu \nu} \right]
\label{lci}
\ee
can be introduced. As is well known, it is metric preserving:
\be
{\stackrel{\circ}{\nabla}}{}_{\rho} g^{\mu \nu} \equiv \partial_\rho
g^{\mu \nu} + {\stackrel{\circ}{\Gamma}}{}^{\mu}{}_{\sigma \rho}
g^{\sigma \nu} + {\stackrel{\circ}{\Gamma}}{}^{\nu}{}_{\sigma \rho}
g^{\mu \sigma} = 0 \; .
\ee
The curvature of the Levi--Civita connection,
\be
{\stackrel{\circ}{R}}{}^{\theta}{}_{\rho \mu \nu} = \partial_\mu
{\stackrel{\circ}{\Gamma}}{}^{\theta}{}_{\rho \nu} +
{\stackrel{\circ}{\Gamma}}{}^{\theta}{}_{\sigma \mu}
\; {\stackrel{\circ}{\Gamma}}{}^{\sigma}{}_{\rho \nu} - (\mu
\leftrightarrow \nu) \; ,
\label{rbola}
\ee
according to general relativity, accounts exactly for the gravitational 
interaction. Owing to the universality of gravitation, which means that
all particles feel ${\stackrel{\circ}{\Gamma}}{}^{\sigma}{}_{\mu \nu}$
the same, it turns out possible to describe the gravitational interaction
by considering a Riemann spacetime with the curvature of the Levi--Civita
connection, in which scalar matter will follow geodesics. This is the
framework of Einstein's general relativity, the gravitational interaction
being mimicked by a geometrization of spacetime. 

A nontrivial tetrad field can also be used to define the zero--curvature
linear Cartan connection 
\be
\Gamma^{\sigma}{}_{\mu \nu} = h_a{}^\sigma \partial_\nu
h^a{}_\mu \; ,
\label{car}
\ee
with respect to which the tetrad is parallel:
\be
{\nabla}_\nu \; h^{a}{}_{\mu} \equiv
\partial_\nu h^{a}{}_{\mu} - \Gamma^{\rho}{}_{\mu \nu} \,
h^{a}{}_{\rho} = 0 \; .
\label{weitz}
\ee
Now, substituting $g^{\mu \nu}$ as given by (\ref{gmn}) into
${\stackrel{\circ}{\Gamma}}{}^{\sigma}{}_{\mu \nu}$, we get the relation
\be
{\Gamma}^{\sigma}{}_{\mu \nu} = {\stackrel{\circ}{\Gamma}}{}^
{\sigma}{}_{\mu \nu} + {K}^{\sigma}{}_{\mu \nu} \; ,
\label{rel}
\ee
where
\be
{K}^{\sigma}{}_{\mu \nu} = \frac{1}{2} \left[
T_{\mu}{}^{\sigma}{}_{\nu} + T_{\nu}{}^{\sigma}{}_{\mu} -
T^{\sigma}{}_{\mu \nu} \right]
\label{conto}
\ee
is the contorsion tensor, with
\be
T^\sigma{}_{\mu \nu} = \Gamma^{\sigma}{}_{\nu \mu} - \Gamma^
{\sigma}{}_{\mu \nu} \; 
\label{tor}
\ee
the torsion of the Cartan connection. If now, analogously to the way
the Riemann spacetime was introduced, we introduce a spacetime
with the same properties of the Cartan connection $\Gamma^{\sigma}{}_{\nu
\mu}$, we end up with a Weitzenb\"ock spacetime,\cite{weitz} a space
presenting torsion, but no curvature. This is the spacetime underlying
the teleparallel description of gravitation. We notice that, if local
Lorentz indices are raised and lowered with the Minkowski metric $\eta^{a
b}$, tensor indices on it will be raised and lowered with the riemannian
metric $g^{\mu \nu}$. Universality of gravitation, in this case, means
that all particles feel $\Gamma^{\sigma}{}_{\nu \mu}$ the same. 

The presence of a nontrivial tetrad field, therefore, induces on
spacetime both a riemannian and a teleparallel structures. The first is
related to the Levi--Civita connection, a connection presenting
curvature, but no torsion. The second is related to the Cartan
connection, a connection presenting torsion, but no curvature. It is
important to notice that there is in this approach no connection
presenting simultaneously non--vanishing curvature and torsion.  

As already remarked, the curvature of the Cartan connection vanishes
identically:
\be
{R}^{\theta}{}_{\rho \mu \nu} = \partial_\mu
{\Gamma}^{\theta}{}_{\rho \nu} + {\Gamma}^{\theta}{}_{\sigma
\mu} \; {\Gamma}^{\sigma}{}_{\rho \nu} - (\mu  \leftrightarrow \nu)
\equiv 0 \; .
\label{r}
\ee
Substituting ${\Gamma}^{\theta}{}_{\mu \nu}$ as given by
Eq.(\ref{rel}), we get
\be
{R}^{\theta}{}_{\rho \mu \nu} =
{\stackrel{\circ}{R}}{}^{\theta}{}_{\rho \mu \nu} +
Q^{\theta}{}_{\rho \mu \nu} \equiv 0 \; ,
\label{eq7}
\ee
where ${\stackrel{\circ}{R}}{}^{\theta}{}_{\rho \mu \nu}$ is the
curvature of the Levi--Civita connection, and
\be
Q^{\theta}{}_{\rho \mu \nu} = D_\mu K^{\theta}{}_{\rho \nu} -
{K}^{\theta}{}_{\sigma \nu} \; K^{\sigma}{}_{\rho \mu} -
(\mu  \leftrightarrow \nu) 
\label{kcur}
\ee
with
\be
D_\mu K^{\theta}{}_{\rho \nu} =
\partial_\mu {K}^{\theta}{}_{\rho \nu} +
{\Gamma}{}^{\theta}{}_{\sigma \mu} \; K^{\sigma}{}_{\rho \nu} -
{\Gamma}{}^{\sigma}{}_{\rho \mu} \; K^{\theta}{}_{\sigma \nu} \; ,
\ee
is a tensor written in terms of the Cartan connection only. Equation
(\ref{eq7}) has an interesting interpretation: the contribution
${\stackrel{\circ}{R}}{}^{\theta}{}_{\rho \mu \nu}$ coming from the
Levi--Civita connection compensates exactly the contribution
$Q^{\theta}{}_{\rho \mu \nu}$ coming from the Cartan connection, yielding
an identically zero Cartan curvature tensor
${R}^{\theta}{}_{\rho \mu \nu}$. This is a constraint satisfied by the
Levi--Civita and Cartan connections, and is the fulcrum of the
equivalence between the riemannian and the teleparallel descriptions of
gravitation.

Now, according to general relativity, the dynamics of the gravitational
field is described by a variational principle with the lagrangian
\be
{\cal L}_g = \frac{\sqrt{-g} \, c^4}{16 \pi G} \; {\stackrel{\circ}{R}}
\; ,
\label{ehl}
\ee
where ${\stackrel{\circ}{R}}= g^{\mu \nu}
{\stackrel{\circ}{R}}{}^{\rho}{}_{\mu \rho \nu}$ is the scalar curvature
of the Levi--Civita connection, and $g={\det}(g_{\mu \nu})$. This
lagrangian, which depends only on the Levi-Civita connection, can be
rewritten in an alternative form depending only on the Cartan connection.
In fact, substituting
${\stackrel{\circ}{R}}$ as obtained from (\ref{eq7}), up to
divergences, we obtain
\be
{\cal L}_g = \frac{h c^4}{16 \pi G} \; \left[\frac{1}{4} \;
T^\rho{}_{\mu \nu} \; T_\rho{}^{\mu \nu} + \frac{1}{2} \;
T^\rho{}_{\mu \nu} \; T^{\nu \mu} {}_\rho - T_{\rho \mu}{}^{\rho}
\; T^{\nu \mu}{}_\nu \right] \; ,
\label{lagr3}
\ee
where $h={\det}(h^a{}_\mu)=\sqrt{-g}$, which is exactly the lagrangian
of a gauge theory for the translation group.\cite{paper} This means
that a translational gauge theory, with a lagrangian quadratic in the
torsion field, is completely equivalent to general relativity, with its
usual lagrangian linear in the scalar curvature. As a
consequence of this equivalence, gravitation will present two
equivalent descriptions, one in terms of a metric geometry, and another
one in which the underlying geometry is that provided by a teleparallel
structure. In what follows, we proceed to describe Maxwell's theory in
the framework of each one of these geometries.

\section{Electromagnetic Field in the Framework of a Riemannian Geometry}

In Minkowski spacetime, the electromagnetic field is described by the
lagrangian density
\be
{\cal L}_{em} = - \frac{1}{4} \, F_{ab} F^{ab} \; ,
\label{l0}
\ee
where
\be
F_{ab} = \partial_a A_b - \partial_b A_a
\label{f0}
\ee
is the Maxwell field-strength tensor. Variation of the corresponding
action in relation to the electromagnetic field $A_a$ yields
\be
\partial_a F^{ab} = 0 \; ,
\label{max0}
\ee
which along with the identity
\be
\partial_a F_{bc} + \partial_c F_{ab} + \partial_b F_{ca} = 0 \; ,
\label{bian0}
\ee
constitutes Maxwell's equations. In the Lorentz gauge, $\partial_a A^a =
0$, and equation (\ref{max0}) can be rewritten as
\be
\partial_c \partial^c A^a = 0 \; .
\label{amax0}
\ee

In the framework of general relativity, the form of Maxwell's theory is
well known. It can be obtained through the application of the so-called
minimal coupling prescription, which amounts to replace
\ba
\eta^{ab} &\rightarrow& g^{\mu \nu} = \eta^{ab} h_{a}{}^{\mu}
h_{b}{}^{\nu} \label{gmc0} \\
{} \nonumber \\
\partial_a &\rightarrow& {\cal D}_\mu = \partial_\mu - \frac{i}{2} \,
{\stackrel{\circ}{\omega}}{}^{ab}{}_{\mu} J_{ab} \; ,
\label{dmc0}
\ea
where ${\cal D}_\mu$ stands for the Fock--Ivanenko derivative
operator,\cite{fock} with
\be
{\stackrel{\circ}{\omega}}{}^{ab}{}_{\mu} = h^{a}{}_{\rho}
{\stackrel{\circ}{\nabla}}{}_{\mu} h^{b \rho}
\label{spin0}
\ee
the spin connection, and $J_{ab}$ an appropriate representation
of the Lorentz group. For the specific case of the (spin 1)
electromagnetic vector field $A^a$,
\be
\left( J_{a b}\right)^c{}_d = i \left( \delta_a{}^c \eta_{b d} -
\delta_b{}^c \eta_{a d} \right) \; ,
\label{vecre}
\ee
and ${\cal D}_{\mu}$ acquires the form
\be
{\cal D}_\mu A^a = \partial_\mu A^a +
{\stackrel{\circ}{\omega}}{}^{a}{}_{b
\mu} \, A^b \; .
\label{fivec1}
\ee

It is important to remark that the Fock--Ivanenko derivative is
concerned {\it only} to the local Lorentz indices. In other words, it
ignores the spacetime tensor
character of the fields being ignored by it.\cite{velt} For example, the
Fock--Ivanenko derivative of the tetrad field is
\be
{\cal D}_\mu h^{a}{}_{\nu} = \partial_\mu h^{a}{}_{\nu} +
{\stackrel{\circ}{\omega}}{}^{a}{}_{b \mu} \, h^{b}{}_{\nu} \; .
\label{fih1}
\ee
Substituting (\ref{spin0}), we get
\be
{\cal D}_\mu h^{a}{}_{\nu} = {\stackrel{\circ}{\Gamma}}{}^{\rho}{}_{\nu
\mu} \, h^{a}{}_{\rho}  \; .
\label{fih2}
\ee
As a consequence, the {\it total} covariant derivative of the tetrad
$h^{a}{}_{\nu}$, that is, a covariant derivative which takes into account
both indices of $h^{a}{}_{\nu}$, vanishes identically:
\be
\partial_\mu h^{a}{}_{\nu} +
{\stackrel{\circ}{\omega}}{}^{a}{}_{b \mu} \, h^{b}{}_{\nu} -
{\stackrel{\circ}{\Gamma}}{}^{\rho}{}_{\nu \mu} \, h^{a}{}_{\rho}  = 0 \; .
\label{fith}
\ee

Now, any Lorentz vector field $A^a$ can be transformed into a spacetime
vector field
$A^\mu$ through
\be
A^\mu = h_{a}{}^{\mu} \, A^a \; ,
\label{aha}
\ee
where $A^\mu$ transforms as a vector under a general spacetime coordinate
transformation. Substituting into equation (\ref{fivec1}), and making use
of (\ref{fih2}), we get
\be
{\cal D}_\mu A^a = h^{a}{}_{\rho} \, {\stackrel{\circ}{\nabla}}{}_{\mu}
A^\rho \; .
\label{fivec3}
\ee  
We see in this way that the Fock--Ivanenko derivative of a Lorentz
vector field $A^a$ reduces to the usual Levi--Civita covariant derivative
of general relativity. This means that the minimal coupling prescription
(\ref{dmc0}) can be restated as
\be
\partial_a \rightarrow {\stackrel{\circ}{\nabla}}{}_{\mu} \equiv
\partial_\mu + {\stackrel{\circ}{\Gamma}}{}_{\mu} \; ,
\ee
which is the form usually presented in the literature.\cite{birrel}
Therefore, in terms of the riemannian structure, the Minkowski lagrangian
(\ref{l0}) acquires the form
\be
{\cal L}_{em} = - \frac{1}{4} \, (-g)^{1/2} \, F_{\mu \nu} F^{\mu \nu} \; ,
\label{l1}
\ee
where
\be
F_{\mu \nu} = {\stackrel{\circ}{\nabla}}{}_{\mu} A_\nu -
{\stackrel{\circ}{\nabla}}{}_{\nu} A_\mu \equiv \partial_{\mu} A_\nu -
\partial_{\nu} A_\mu \; ,
\label{f1}
\ee
the connection terms canceling due to the symmetry of the Levi--Civita
connection in the last two indices. The corresponding Maxwell's equation is
\be
{\stackrel{\circ}{\nabla}}{}_{\mu} F^{\mu \nu} = 0 \; ,
\label{max1}
\ee
or equivalently, assuming the covariant Lorentz gauge
${\stackrel{\circ}{\nabla}}{}_{\mu} A^\mu = 0$,
\be
{\stackrel{\circ}{\nabla}}{}_{\mu} {\stackrel{\circ}{\nabla}}{}^{\mu}
A_{\nu} - {\stackrel{\circ}{R}}{}^{\mu}{}_{\nu} A_{\mu} = 0 \; .
\label{amax1}
\ee
Analogously, the Bianchi identity (\ref{bian0}) can be shown to assume
the form
\be
\partial_\mu F_{\nu \sigma} + \partial_\sigma F_{\mu \nu} + \partial_\nu
F_{\sigma \mu} = 0 \; .
\label{bian1}
\ee
We notice in passing that the presence of curvature does not spoil the gauge
invariance of Maxwell theory. 
  
Let us now consider the system formed by gravitational plus
electromagnetic fields, represented by the lagrangian ${\cal L} = {\cal
L}_{g} + {\cal L}_{em}$, with ${\cal L}_{g}$ given by (\ref{ehl}), and
${\cal L}_{em}$ by (\ref{l1}). Variation of the corresponding action in
relation to the metric $g^{\mu \nu}$ yields Einstein's field equation
\be
{\stackrel{\circ}{R}}{}_{\mu \nu} - \frac{1}{2} \; 
g_{\mu \nu} \; {\stackrel{\circ}{R}} = \frac{8 \pi G}{c^4} \,
{\cal T}_{\mu \nu} \; ,
\label{einstein}
\ee
where
\be
{\cal T}_{\mu \nu} = - \frac{2}{\sqrt{-g}} \, \frac{\delta {\cal
L}_{em}}{\delta g^{\mu \nu}} = \frac{1}{4} \left[ F_{\mu}{}^{\rho}
F_{\nu \rho} - \frac{1}{4} g_{\mu \nu} F_{\rho \sigma} F^{\rho \sigma}
\right]
\ee
is the energy--momentum tensor of the electromagnetic field. From these
considerations, therefore, we can conclude that, in the framework
of general relativity, the elec\-tro\-mag\-netic field is able to
feel as well as to produce curvature. 

\section{Electromagnetic Field in the Framework of a Teleparallel Geometry}

In the preceding section, we have described Maxwell's theory
in terms of the riemannian structure of spacetime. Now, we obtain
Maxwell's theory in terms of the
teleparallel structure of spacetime. To start with, we notice that, from
(\ref{weitz}) and (\ref{rel}), we get
\be
{\stackrel{\circ}{\nabla}}{}_{\mu} h^{b \rho} = - K^{\rho \nu}{}_{\mu}
h^{b}{}_{\nu} \; .
\label{spin1}
\ee
Thus, in terms of magnitudes related to the teleparallel structure, the
spin connection (\ref{spin0}) is written as
\be
{\stackrel{\circ}{\omega}}{}^{ab}{}_{\mu} = - h^{a}{}_{\rho}
K^{\rho \nu}{}_{\mu} h^{b}{}_{\nu} \; .
\label{spin2}
\ee
We remark that this equation could also be obtained directly from relation
(\ref{rel}) by transforming spacetime into algebra indices. It is important
to remember, however, that connections transform not
covariantly under a basis transformation.\cite{livro} For example,
\[
{\stackrel{\circ}{\omega}}{}^{a}{}_{b\mu} = h^{a}{}_{\rho}
{\stackrel{\circ}{\Gamma}}{}^{\rho}{}_{\lambda \mu} h_{b}{}^{\lambda} +
h^{a}{}_{\rho} \partial_\mu h_{b}{}^{\rho} \; ,
\]
and, denoting by ${\omega}{}^{a}{}_{b\mu}$ the transformed Cartan connection,
we have also
\[
{\omega}{}^{a}{}_{b\mu} = h^{a}{}_{\rho}
{\Gamma}{}^{\rho}{}_{\lambda \mu} h_{b}{}^{\lambda} +
h^{a}{}_{\rho} \partial_\mu h_{b}{}^{\rho} \equiv 0 \; ,
\]
the vanishing of ${\omega}{}^{a}{}_{b\mu}$ coming from the absolute
parallelism condition (\ref{weitz}). We see in this way that
(\ref{spin2}) has implicitly the zero--connection
${\omega}{}^{a}{}_{b\mu}$ on its right--hand side, which explains why a
connection can apparently be equal to a tensor. The teleparallel version
of the minimal coupling prescription, therefore, can be stated as
\ba
\eta^{ab} &\rightarrow& g^{\mu \nu} = \eta^{ab} h_{a}{}^{\mu}
h_{b}{}^{\nu} \; ,
\label{gmc1} \\
{} \nonumber \\
\partial_a &\rightarrow& {\cal D}_{\mu} = \partial_\mu + \frac{i}{2}
\, h^{a}{}_{\rho} K^{\rho \nu}{}_{\mu} h^{b}{}_{\nu} \, J_{ab} \; ,
\label{dmc1}
\ea
with ${\cal D}_{\mu}$ standing now for the teleparallel version of the
Fock--Ivanenko derivative operator. In the specific case of the
electromagnetic vector field $A^a$,
$J_{ab}$ is given by (\ref{vecre}), and we get
\be
{\cal D}_\mu A^c = \partial_\mu A^c - h^c{}_{\rho} K^{\rho}{}_{\nu \mu} 
\, h_{d}{}^{\nu} \, A^d \; .
\label{dac}
\ee
This is the teleparallel version of the Fock--Ivanenko derivative of a
vector field $A^c$. To obtain the corresponding covariant derivative of
the spacetime vector field $A^\nu$, we substitute $A^d=h^{d}{}_{\nu} 
A^\nu$ in the right-hand side. The result is
\be
{\cal D}_\mu A^c = h^c{}_{\rho} \left[ \partial_\mu A^\rho -
K^{\rho}{}_{\nu
\mu} \, A^\nu \right] + A^\lambda \partial_\mu h^c{}_{\lambda} \; .
\label{dac1}
\ee
Now, from equation (\ref{car}) we see that
\be
\partial_\mu h^c{}_{\lambda} = h^c{}_{\rho} \Gamma^{\rho}{}_{\lambda \mu} 
\; .
\label{dh}
\ee
Consequently, equation (\ref{dac1}) acquires the form
\be
{\cal D}_\mu A^c = h^c{}_{\rho}  D_\mu A^\rho \; ,
\label{dac2}
\ee
where
\be
D_\mu A^\rho = \nabla_\mu A^\rho - K^{\rho}{}_{\nu \mu} \, A^\nu
\label{telcode}
\ee
is the teleparallel version of the covariant derivative, with
\be
\nabla_\mu A^\rho = \partial_\mu A^\rho + \Gamma^{\rho}{}_{\nu \mu} \,
A^\nu
\label{carcode}
\ee
the Cartan covariant derivative. This means that the teleparallel version of
the minimal coupling prescription (\ref{dmc1}) can be restated as
\be
\partial_a \rightarrow D_\mu \equiv \partial_\mu +
\Gamma_\mu - K_\mu \; . 
\label{dmc2}
\ee

In terms of the teleparallel structure, therefore, the free
lagrangian (\ref{l0}) becomes
\be
{\cal L}_{em} = - \frac{h}{4} \, F_{\mu \nu} F^{\mu \nu}
\label{l2}
\ee
where now
\be
 F_{\mu \nu} = D_\mu A_\nu - D_\nu A_\mu \; .
\label{f2}
\ee
Using the explicit form of $D_\mu$ and the definitions of torsion and
contorsion tensors, it is an easy task to verify that
\be
 F_{\mu \nu} = \partial_\mu A_\nu - \partial_\nu A_\mu \; ,
\label{f3}
\ee
which is a gauge invariant tensor. Variation of the corresponding action
in relation to the electromagnetic field $A^\mu$ yields the teleparallel
version of the first pair of Maxwell's equation:
\be
D_\mu F^{\mu \nu} = 0 \; .
\label{max2}
\ee
Equivalently, assuming the teleparallel Lorentz gauge $D_\mu A^\mu = 0$,
and using the commutation relation
\be
\left[ D_\mu, D_\nu \right] A^\mu = - Q_{\mu \nu} \, A^\mu \; ,
\ee
where $Q_{\mu \nu} = Q^{\rho}{}_{\mu \rho \nu}$, with
$Q^{\rho}{}_{\mu \sigma \nu}$ given by equation (\ref{kcur}), we obtain
\be
D_{\mu} D^{\mu} A_{\nu} + Q^{\mu}{}_{\nu} A_{\mu} = 0 \; .
\label{amax2}
\ee
On the other hand, by using the same coupling prescription in the
Bianchi identity (\ref{bian0}), the teleparallel version of
the second pair of Maxwell's equation becomes
\be
\partial_\mu F_{\nu \sigma} + \partial_\sigma F_{\mu \nu} + \partial_\nu
F_{\sigma \mu} = 0 \; .
\label{bian2}
\ee
We see in this way that, in the context of the teleparallel equivalent
of general relativity, the electromagnetic field is able to
couple to torsion, and that this coupling does
not violate the U(1) gauge invariance of Maxwell's theory.
Furthermore, using the relation (\ref{rel}), it is easy to verify that
the teleparallel version of Maxwell's equations, which are equations
written in terms of the Cartan connection only, are completely
equivalent to the usual Maxwell's equations in a riemannian
background, which are equations written in terms of the Levi--Civita
connection only.

Let us now consider the system formed by gravitational plus
electromagnetic fields, represented by the lagrangian ${\cal L} = {\cal
L}_{g} + {\cal L}_{em}$, with ${\cal L}_{g}$ given by (\ref{lagr3}), and
${\cal L}_{em}$ by  (\ref{l2}). Variation of the corresponding action in
relation to the tetrad field yields the teleparallel version of
Einstein's equations,
\be
\partial_\rho \; S_{\mu}{}^{\nu \rho} - 
\frac{4 \pi G}{c^4} \; t_{\mu}{}^{\nu} = \frac{4 \pi G}{c^4} \;
{\cal T}_{\mu}{}^{\nu} \; ,
\label{ym}
\ee
where $S_{\mu}{}^{\nu \rho}= - S_{\mu}{}^{\rho \nu}$ is given by 
\[
S_{\mu}{}^{\nu \rho} = \frac{1}{4} \; \left(T_{\mu}{}^{\nu \rho} +
T^{\nu}{}_{\mu}{}^{\rho} - T^{\rho}{}_{\mu}{}^{\nu}\right) - \frac{1}{2}
\;
\left(\delta_{\mu}{}^{\rho} \; T_{\theta}{}^{\nu \theta} -
\delta_{\mu}{}^{\nu} \; T_{\theta}{}^{\rho \theta} \right) \; ,
\]
$t_{\mu}{}^{\nu}$ is the energy--momentum (pseudo) tensor of the
gravitational field, and
\[
{\cal T}_{\mu}{}^{\nu} = h^{\alpha}{}_{\mu} \left( - \frac{1}{h} \,
\frac{\delta {\cal L}_{em}}{\delta h^{\alpha}{}_{\nu}} \right)  
\]
is the energy--momentum tensor of the electromagnetic field. In the
teleparallel description of gravitation, therefore, besides coupling to
torsion in a gauge covariant way, the electromagnetic field, through
its energy--momentum tensor, is also able
to produce torsion, a point which is not in agreement with the usual
belief that only a spin distribution could be the source of torsion.

\section{Final Comments}

A nontrivial tetrad field induces on spacetime both a 
riemannian and a teleparallel structures.
General relativity, a theory formulated in terms of the riemannian
structure, is known to be completely equivalent to a gauge theory for the 
translation group, whose underlying geometry is that provided by the
teleparallel structure. We should remark that, as no connection
presenting simultaneously non--vanishing
curvature and torsion is present in this formalism, no
Riemann--Cartan spacetime enters the description of the gravitational
interaction.

Using the above approach, we have studied in this paper the
coupling of the electromagnetic field with gravitation, the latter being
described in terms of the teleparallel
structure of spacetime. In terms of the riemannian structure, that is, in
the framework of general relativity, the form of Maxwell's theory is well
known, and it has been reviewed here for completeness reason. In terms of
the teleparallel structure, we have shown that, provided
an appropriate coupling prescription is used, the
electromagnetic field is able not only to couple to torsion, but also,
through its energy--momentum tensor, to produce torsion. Furthermore,
we have shown that the coupling of the electromagnetic field with torsion
does not violate the local gauge invariance of Maxwell's theory.

The crucial point of our approach is the introduction of the teleparallel
coupling prescription (\ref{dmc2}), which is a natural consequence of the
assumed equivalence between general relativity and a gauge theory for the
translation group, and which we believe accounts correctly for the
coupling between Maxwell's field and torsion. It is not {\it minimal} in the
usual sense as such a name is currently reserved for couplings of the form
\[
\partial_a \rightarrow \nabla_\mu \equiv \partial_\mu - \Gamma_\mu \; .
\]
On the other hand, it is just what is needed so that it turns out to be
equivalent to the minimal coupling of the riemannian description.
In fact, anyone of the versions of Maxwell's
equations, the teleparallel or the metric one, can be obtained from
each other by using the relation (\ref{rel}). Finally, it is worth mentioning
that the same coupling prescription has been used to study the coupling of a
scalar field with both curvature and torsion tensors, showing that this
formalism can be consistently applied to other fields as well.\cite{paper2}

\section*{Acknowledgments}

The authors would like to thank R. Aldrovandi for useful discussions. They would
also like to thank FAPESP--Brazil and CNPq--Brazil, for financial support.

\end{document}